\documentclass[12pt]{iopart}

\usepackage{xcolor}
\usepackage{amssymb,amsfonts,bm}
\usepackage{url}
\usepackage{graphicx}
\usepackage[labelfont=bf, font=small]{caption}
\usepackage{harvard}  
\citationmode{abbr}

\newcommand{\x}{\mathbf{x}}
\newcommand{\rvec}{\mathbf{r}}
\newcommand{\kvec}{\mathbf{k}}  

\usepackage[normalem]{ulem}

\begin{document}

\title[Pulse-echo ultrasound attenuation tomography]{Pulse-echo ultrasound attenuation tomography}

\author{Naiara {Korta Martiartu}$^*$, Parisa {Salemi Yolgunlu}, Martin Frenz, and Michael Jaeger}
\address{Institute of Applied Physics, University of Bern, Sidlerstrasse 5, 3012 Bern, Switzerland}
\address{$^*$Author to whom any correspondence should be addressed.}
\ead{naiara.korta@unibe.ch}


\begin{abstract}
\textit{Objective.}~We present the first fully two-dimensional attenuation imaging technique developed for pulse-echo ultrasound systems. Unlike state-of-the-art techniques, which use line-by-line acquisitions, our method uses steered emissions to constrain attenuation values at each location with multiple crossing wave paths, essential to resolve the spatial variations of this tissue property. \textit{Approach.}~At every location, we compute normalized cross-correlations between the beamformed images that are obtained from emissions at different steering angles. We demonstrate that their log-amplitudes provide the changes between attenuation-induced amplitude losses undergone by the different incident waves. This allows us to formulate a linear tomographic problem, which we efficiently solve via a Tikhonov-regularized least-squares approach. \textit{Main results.}~The performance of our tomography technique is first validated in numerical examples and then experimentally demonstrated in custom-made tissue-mimicking phantoms with
inclusions of varying size, echogenicity, and attenuation. We show that this technique is particularly good at resolving lateral variations in tissue attenuation and remains accurate in media with varying echogenicity. \textit{Significance.}~Based on a similar principle, this method can be easily combined with computed ultrasound tomography in echo mode (CUTE) for speed-of-sound imaging, paving the way towards a multi-modal ultrasound tomography framework characterizing multiple acoustic tissue properties simultaneously.
\end{abstract}

%
\vspace{2pc}

\submitto{\PMB}
%
\maketitle
%
%

\section{Introduction}

Ultrasound waves propagating through tissues undergo energy losses mainly due to absorption and scattering~\cite{SZABO2014}. This attenuation significantly differs between tissue types and can reveal disease-related changes in tissue composition and structure~\cite{DUCK1990}. For example, the controlled attenuation parameter (CAP), which estimates the average ultrasound attenuation in the liver, is a widely used noninvasive tool to assess diffuse hepatic steatosis, i.e., fat accumulation in the liver~\cite{eddowes2019accuracy,Cao_2022_eClinicalMedicine,An_2022_HepatComm}. Despite being a powerful diagnostic technique, CAP is designed to provide a single value corresponding to a small volume within the liver~\cite{SASSO2010} and cannot thus characterize the inherent tissue heterogeneity. This hinders its extension to clinical applications involving focal lesions or other organs, such as the breast, where attenuation has proven relevant for differentiating malignant and benign lesions~\cite{Calderon1976,DASTOUS1986,Nam2013}. 

Imaging techniques that quantify the spatial distribution of ultrasound attenuation in tissue can offer a versatile tool to overcome the limitations of CAP. Several approaches have been suggested using either time-domain~\cite{Ghoshal2012} or frequency-domain formulations~\cite{KANAYAMA2013,Pawlicki2013,Colia2018,Brandner21,KIM2008}. Although computationally more demanding, the latter are more common as they provide efficient ways to compensate for system- and transducer-dependent effects. Frequency-domain techniques retrieve the attenuation in tissue by analyzing the spectral information of backscattered radio-frequency signals. They are classified into three types depending on the information they exploit~\cite{Labyed2011}, which can be either frequency-dependent amplitude variations (spectral difference and spectral-log difference methods), the downshift of the center frequency (spectral shift methods), or a combination of the two (hybrid methods). Independent of the type or domain, nearly all these techniques consider conventional line-by-line scanning to relate data variations with depth to attenuation. That is, they use simplified one-dimensional formulations where ultrasound waves propagate only in the axial direction, limiting the amount of data used to constrain the attenuation at each location and, thus, limiting the spatial resolution and accuracy of reconstructed images~\cite{Labyed2011,Samimi2017}.

To account for the two-dimensional (2-D) nature of attenuation in tissue, \citeasnoun{Colia2018} reformulated the spectral-log difference technique as a regularized inverse problem. Instead of estimating the attenuation at each depth independently, their method includes 2-D spatial constraints via regularization. They show that this can substantially improve the trade-off between the spatial resolution and accuracy of state-of-the-art techniques; however, regularization strategies rely on subjective choices rather than actual 2-D data constraints, which would require interrogating each tissue location with ultrasound waves propagating along multiple directions. An attempt in this direction was suggested by \citeasnoun{TREECE2007} to correct shadowing and enhancement artifacts in B-mode images. They developed a method that estimates the lateral variations in the cumulative attenuation from pairs of beams steered at opposite angles. Although successful at restoring B-mode images, their technique cannot characterize the local attenuation in tissue since this information is not actually required for artifact correction. 

In this work, we present a fundamentally new 2-D approach to quantify the spatial distribution of tissue attenuation using pulse-echo ultrasound systems. We suggest using a set of steered emissions in order to measure the changes between the amplitudes of echoes that are detected using different steering angles. We can relate these to the attenuation-induced amplitude losses undergone by the incident waves, allowing us to formulate a tomographic technique to reconstruct tissue attenuation. Computed ultrasound tomography in echo mode (CUTE) uses a similar approach to relate tissue speed of sound to the echo phase shifts observed when probing tissue at different angles~\cite{jaeger2015computed,stahli2019forward,jaeger2022pulseecho} and has demonstrated unprecedented spatial and contrast resolution in both tissue-mimicking phantoms~\cite{stahli2020bayesian} and in-vivo~\cite{NatureCUTE2022}. In the following, we first introduce the theoretical foundations of the proposed attenuation imaging technique in~\sref{sec:theory} and discuss its practical implementation in~\sref{sec:implementation}. We then validate this approach using numerical wave propagation simulations in~\sref{sec:num_results} and finally demonstrate its experimental feasibility in~\sref{sec:exp_results} using calibrated tissue-mimicking phantoms with varying acoustic properties.

\section{Theory}
\label{sec:theory}

\subsection{Beamformed ultrasound images}
\label{sec:FDbeamforming}

In pulse-echo ultrasound, we use a transducer array to emit pressure waves through the tissue and acquire the backscattered wavefield at each transducer element. If we neglect multiple scattering and denote $\chi(\x)$ the scattering function of the medium defined in $\Omega \subset \mathbb{R}^{2}$, we can generally express the backscattered wavefield sensed at $\x_r$ in the temporal-frequency domain as
\begin{equation}
p(\x_r,\omega) =  \int_\Omega
 p_e(\x, \omega) \chi(\x) G(\x_r,\x,\omega)
d\x,
\label{eq:representation}
\end{equation}
where $\omega = 2\pi f$ is the angular frequency of the waves, $p_e(\x, \omega)$ refers to the emitted pressure field, and $G(\x_r,\x,\omega)$ is the Green's function, i.e., the impulse response of the medium between locations $\x$ and $\x_r$~\cite{Jensen91}. In this study, we consider insonifying the tissue with steered plane waves, i.e.,
\begin{equation}
p_e(\x, \omega) =  A(\omega)\exp(j\kvec\cdot\x),
 \label{eq:PlaneWave}
\end{equation}
where $A$ denotes the amplitude of the plane wave, and the wave number vector $\kvec$ defines its propagation direction. Since tissue is attenuating, this vector is considered complex, i.e., $\kvec = \kvec^r + j \kvec^i$. The real part $k^r = \omega/c(\omega)$ encodes the wave phase velocity $c$, whereas the imaginary part $k^i = \alpha(\omega)$ describes the ultrasound attenuation (in units of Np/m) exhibiting the frequency power law relation $\alpha(\omega) = \alpha_0\omega^y$ in tissues~\cite{SZABO2014}. Here $\alpha_0$ is the attenuation coefficient, and $y$ is the power law exponent typically ranging from 1 to 2. In the following, we omit the dependency on $\omega$ for a condensed notation.

We can generate an ultrasound image $I(\rvec)$ per plane-wave emission by applying delay-and-sum beamforming on recorded signals $p(\x_r)$. This process typically assumes a lossless homogeneous medium (indicated with the subscript $0$) to focus the received energy into an arbitrary focal point $\rvec$ via the operation
\begin{equation}
I(\rvec) = \int_\Omega A \exp\left[j(\kvec\cdot\x - \kvec_0\cdot\rvec)\right] \chi(\x)  H(\x, \rvec) d\x,
 \label{eq:Image}
\end{equation}
where
\begin{equation}
H(\x, \rvec) = \sum_{x_r} G(\x_r,\x) G^*_0(\x_r,\rvec)
\label{eq:receivePSF}
\end{equation}
is the receive point-spread function~\cite{Lambert2020Reflection}.
Here $*$ indicates the complex conjugate operation, and the summation is performed over all receiving elements. Note that $\kvec_0$ is related to the beamforming process and is thus a real vector. Although we typically integrate the equation \eref{eq:Image} over the frequency range of ultrasound pulses to obtain high-resolution images, the next subsection considers the monochromatic formulation for simplicity.

\subsection{Relationship between attenuation and cross-correlations of beamformed images}
\label{sec:xcorrtheory}

Assume we emit two plane waves steered at directions $\kvec_1$ and $\kvec_2$ sequentially and reconstruct the corresponding images $I_1(\rvec)$ and $I_2(\rvec)$, respectively. As a first approximation, we consider tissue as a diffuse scattering medium, meaning that scatterers are spatially uncorrelated and satisfy $\langle
\chi^*(\x_2)\chi(\x_1) \rangle = \langle|\chi|^2\rangle\delta(\x_1 - \x_2)$, where $\langle \cdot \rangle$ denotes an ensemble average and $\delta$ is the Dirac distribution~\cite{Mallart91,Lambert2020Reflection}. Then, the ensemble-averaged cross-correlation between the two images $C_{12}(\rvec) := \langle I^*_2(\rvec) I_1(\rvec) \rangle$ is given by
\begin{equation}
C_{12}(\rvec) = \int_\Omega \langle|\chi|^2\rangle A_1A_2 \exp\left[j(\delta\kvec\cdot\x - \delta\kvec_0\cdot\rvec)\right]  | H(\x,\rvec)|^2 d\x,
 \label{eq:Crosscorrelation}
\end{equation}
with $\delta\kvec = \kvec_1 - \kvec^*_2$ and $\delta\kvec_0 = \kvec_{1,0} - \kvec_{2,0}$. $A_i$ is the amplitude of the emitted plane wave along $\kvec_i$ direction.

The point-spread functions in equation \eref{eq:Image} restrict the range of locations significantly contributing to any arbitrary image position $\rvec_0$. Thus, without loss of generality, we can focus our analysis of $C_{12}$ to a subvolume $\Omega_0 \subset \Omega$ centered around $\rvec_0$, where $\x  = \rvec_0 + \bm \xi \in \Omega_0$. In this case, $C_{12}$ at $\rvec_0$ reduces to
\begin{equation}
C_{12} (\rvec_0) \approx  \langle|\chi|^2\rangle A_1 A_2 \exp[j(\delta\kvec - \delta\kvec_0)\cdot\rvec_0] \int_{\Omega_0} | H(\bm\xi,\rvec_0)|^2 \exp[j\delta\kvec\cdot\bm\xi] d\bm\xi.
 \label{eq:Crosscorrelation_approx}
\end{equation}

When the difference between the plane-wave steering angles $\Delta\varphi$ is sufficiently small to satisfy
 \begin{equation}
\left|(\kvec_1 - \kvec^*_2)\cdot\bm{\xi}\right| \ll 1 
\quad \Rightarrow \quad ||\bm{\xi}|| \ll
\frac{\lambda}{2\pi\Delta\varphi}, \quad \forall \bm{\xi}\in\Omega_0,
\label{condition_plane_nearfield2}
\end{equation}
where the loss per wavelength is assumed small, i.e., $k^i \ll k^r$~\cite{Szabo94}, the exponential inside the integral in equation~\eref{eq:Crosscorrelation_approx} approximates to 1, and the cross-correlation becomes
\begin{equation}
C_{12} (\rvec_0) \approx B(\rvec_0) \exp[j(\kvec^r_1  - \kvec^r_{1,0} - \kvec^r_2 + \kvec^r_{2,0})\cdot\rvec_0] \exp[-(\kvec^i_1 + \kvec^i_2)\cdot\rvec_0],
\label{crosscorrelation_final}
\end{equation}
with 
\begin{equation}
 B(\rvec_0) = \langle|\chi|^2\rangle A_1 A_2 \int_{\Omega_0} | H(\bm\xi,\rvec_0)|^2 d\bm\xi.
\label{crosscorrelation_final2}
\end{equation}
Thus, in a first approximation, when plane-wave emissions satisfy equation \eref{condition_plane_nearfield2}, only the propagation time differences related to the incident waves contribute to the phase of the cross-correlation. This approximation led to the first development of CUTE for estimating the speed-of-sound distribution in tissue~\cite{jaeger2015computed}, and we propose a similar approach for attenuation imaging in this work. In order to isolate the attenuation-induced amplitude terms from equation \eref{crosscorrelation_final}, we suggest dividing $C_{12}$ with the auto-correlation 
\begin{equation}
C_{11} (\rvec_0) \approx\langle|\chi|^2\rangle A^2_1 \exp[-2\kvec^i_1\cdot\rvec_0]\int_{\Omega_0} | H(\bm\xi,\rvec_0)|^2 d\bm\xi.
\label{autocorrelation}
\end{equation}
Then, the log-amplitude of the normalized cross-correlation is
\begin{equation}
(\kvec^i_2 - \kvec^i_1)\cdot\rvec_0 \approx -\log\left|\frac{C_{12} (\rvec_0)}{C_{11} (\rvec_0)}\right|,
\label{forward}
\end{equation}
where we assumed $A_1 \approx A_2$ following the requirement of small steering angle differences previously discussed. Thus, the normalized cross-correlation operation is equivalent to placing a virtual receiver at $\rvec_0$ that measures the changes between attenuation-induced amplitude losses undergone by the incident waves. Equation~\eref{forward} establishes the forward problem of the proposed attenuation imaging technique in this work. Note that if we use $C_{22}$ for the normalization instead, only a sign change occurs in equation~\eref{forward}; so, we can alternatively express the forward problem as
\begin{equation}
(\kvec^i_2 - \kvec^i_1)\cdot\rvec_0 \approx -\frac{1}{2}\log\left|\frac{C_{12} (\rvec_0)}{C_{11} (\rvec_0)}\right| + \frac{1}{2}\log\left|\frac{C_{12} (\rvec_0)}{C_{22} (\rvec_0)}\right|,
\label{forward_sym}
\end{equation}
which may be useful to reduce the observational noise via the averaging of both log-amplitude terms. 

The validity of this forward problem is subject to condition \eref{condition_plane_nearfield2}, which depends on the size of point-spread functions in equation \eref{eq:Image} and, thus, on the focusing quality of beamformed images. For instance, for steering angle differences of $\Delta\varphi = 2.5^\circ$ (see next section), we would require a $||\bm{\xi}||$ approximately smaller than $4 \lambda$, which is considerably larger than the size of diffraction-limited point-spread functions arising when aberrations are not present~\cite{Chen2020}. This condition may not hold when the assumed and true impulse responses in equation \eref{eq:Image} differ substantially. In this case, the normalization operation cannot accurately isolate attenuation-induced amplitude losses of the incident  waves, and we may expect an increased noise level in the log-amplitude measurements. We will discuss this point further in \sref{sec:sensiSoS}. 

\subsection{Inversion}
\label{sec:inverse}

In attenuation imaging, it is common to consider ultrasound waves propagating as straight rays. Under this simplification, the term on the right-hand side of equation~\eref{forward} becomes
\begin{equation}
(\kvec^i_2 - \kvec^i_1)\cdot\rvec_0 \approx \int_{L_2(\rvec_0)} \alpha(\rvec) dl - \int_{L_1(\rvec_0)} \alpha(\rvec) dl,
\label{forward_right}
\end{equation}
where $L_i(\rvec_0)$ refers to the ray path connecting the transducer array and position $\rvec_0$ along direction $\kvec_i$, and $dl$ is the differential arc length along this ray. Upon inserting equation~\eref{forward_right} into equation~\eref{forward} or~\eref{forward_sym}, we can express the forward problem in matrix notation as
\begin{equation}
\mathbf{d} = \mathbf{F} \mathbf{m},
\label{eq:forwardMat}
\end{equation}
where $\mathbf{d} \in \mathbb{R}^{N}$ is a vector containing normalized cross-correlation log-amplitude measurements for each tissue location and pair of incident  waves, $\mathbf{m} \in \mathbb{R}^{M}$ contains the attenuation values $\alpha$ at each location of the tissue (discretized here using a rectilinear grid), and $\mathbf{F} \in \mathbb{R}^{N \times M}$ is the forward operator whose rows contain differences between pairs of ray paths for each measurement location. 

Due to the limited data coverage, the forward operator $\mathbf{F}$ is ill-conditioned, and the inverse problem requires regularization. To ensure close-to-real-time solutions critical to medical ultrasound, we use first-order Tikohonov regularization, for which the inversion has the closed-form solution
\begin{equation}
\hat{\mathbf{ m}} = \left(\mathbf{F}^\mathsf{T}\mathbf{F} + \lambda \mathbf{D}^\mathsf{T} \mathbf{D} \right)^{-1} \mathbf{F}^\mathsf{T} \mathbf{ d}.
\label{eq:inversion}
\end{equation}
Here, the superscript $\mathsf{T}$ stands for the matrix transpose operation, $\mathbf{D} = [\mathbf{D_x}~\mathbf{D_z}]^\mathsf{T}$ contains the first-order finite-difference operators along the lateral (x) and axial (z) directions penalizing non-smooth solutions, and $\lambda$ is the regularization parameter, which we optimize using the L-curve method~\cite{Hansen93}. Note that equation \eref{eq:inversion} assumes normally distributed and uncorrelated noise, equal for all observations. As observed in pulse-echo speed-of-sound tomography, reconstructions can benefit from anisotropic regularization, particularly in layered media~\cite{Sanabria_2018,stahli2019forward}. This can be implemented by reformulating $\lambda \mathbf{D}^\mathsf{T} \mathbf{D}$ as $\lambda_x \mathbf{D_x}^\mathsf{T} \mathbf{D_x} + \lambda_z \mathbf{D_z}^\mathsf{T} \mathbf{D_z}$, with different regularization strengths along each direction. The L-curve approach cannot optimize more than one regularization parameter without additional constraints. Thus, we keep a fixed $\lambda_x/\lambda_z$ ratio to apply this approach for anisotropic regularization.

\subsection{Uncertainty assessment}

Uncertainties of attenuation estimates $\hat{\mathbf{m}}$ in equation~\eref{eq:inversion} are described by means of the posterior covariance matrix $\bm{\Gamma}_\mathrm{post} := \left(\mathbf{F}^\mathsf{T}\mathbf{F} + \lambda \mathbf{D}^\mathsf{T} \mathbf{D} \right)^{-1}$.
For example, its diagonal elements provide the variances of $\hat{\mathbf{m}}$ and can be used to visualize how reliable our attenuation estimates are in each location.
We also consider here the resolution matrix $\mathbf{R}$, which relates the true medium with the estimated one. Assuming that our forward model is perfectly accurate, we can relate the observations to a true tissue model satisfying $\mathbf{d} = \mathbf{F} \mathbf{m}_\mathrm{true}$ and replace this in equation~\eref{eq:inversion} to obtain
\begin{equation}
\hat{\mathbf{ m}} = \bm{\Gamma}_\mathrm{post} \mathbf{F}^\mathsf{T}  \mathbf{F} \mathbf{m}_\mathrm{true} = \mathbf{R} \mathbf{m}_\mathrm{true}.
\label{eq:inversion_uncertainty}
\end{equation}
The columns of $\mathbf{R}$
provide the relationship between reconstructed attenuation values at different locations, i.e., the tomographic point-spread functions~\cite{KortaMartiartu2020}. Note that since our problem is linear, both $\bm{\Gamma}_\mathrm{post}$ and $\mathbf{R}$ are independent of $\mathbf{m}$ and can be precomputed to provide fast images and uncertainty estimates in practice.

\section{Practical implementation}
\label{sec:implementation}

The proposed attenuation tomography technique is conceptually similar to the speed-of-sound imaging technique CUTE. Its practical implementation thus closely follows the one typically used in CUTE. We describe its most important aspects here and refer the reader to~\citeasnoun{stahli2019forward} for a more detailed justification of the choice of acquisition and processing parameters.

\subsection{Ultrasound data acquisition and beamforming}

While the theoretical derivations presented so far are valid regardless of the ultrasound probe type, this work focuses on linear transducer arrays for simplicity. We consider an acquisition sequence consisting of steered plane-wave emissions at angles $\varphi$ ranging from -27.5$^\circ$ to 27.5$^\circ$, with an angular step of 0.5$^\circ$. 
Typical linear probes have an inter-element pitch close to the wavelength; thus, this angle range is limited by grating lobes becoming dominant at larger angles. The small angular step is required for the coherent compounding described in the next subsection. For each emission, we collect the backscattered signals $p(\x_r, t)$ arriving at each transducer element and transform them into the analytic signals $\hat{p}(\x_r, t)$ using the Hilbert transform. Section~\ref{sec:FDbeamforming} described the beamforming process in the temporal-frequency domain to simplify the theoretical derivations. In practice, however, we implement the delay-and-sum algorithm in the time domain and reconstruct the image $I_\varphi(\rvec)$ corresponding to the emission along $\varphi$ as
\begin{equation}
I_\varphi(\rvec) = \sum_{\x_r} a(\rvec,\x_r) \hat{p}(\x_r, t_0(\rvec))
\end{equation}
using the focusing law 
\begin{equation}
t_0(\rvec) = \left[x\sin\varphi + z\cos\varphi + \sqrt{(x-x_r)^2 + (z-z_r)^2}\right]/c_0,
\end{equation}
where $\rvec = (x, z)$, $c_0$ refers to the assumed tissue speed of sound, and $a$ is the apodization factor limiting the receiving angular aperture. Note that this factor was omitted in~\sref{sec:FDbeamforming} for simplicity, although it could have been included in equation~\eref{eq:receivePSF}. The apodization factor does not explicitly affect the derivation of the forward problem~\eref{forward}. However, it shapes the receive point-spread function and may therefore increase the size of $\Omega_0$, affecting the fulfillment of condition~\eref{condition_plane_nearfield2}. Here, we use a factor $a$ that limits the receiving angular aperture to $\pm30^\circ$, which was empirically found to provide a good trade-off between the focusing quality and observational noise. Furthermore, we use a rectilinear grid for reconstructing the images $I_\varphi(\rvec)$, with the lateral and axial grid spacing equal to the element pitch and $c_0\Delta t/2$, respectively, where $\Delta t$ refers to the sampling period of~$\hat{p}(\x_r, t)$.

\subsection{Normalized cross-correlation log-amplitude measurements}
\label{sec:measurements}
\begin{figure}[t]
\includegraphics[width=1\textwidth]{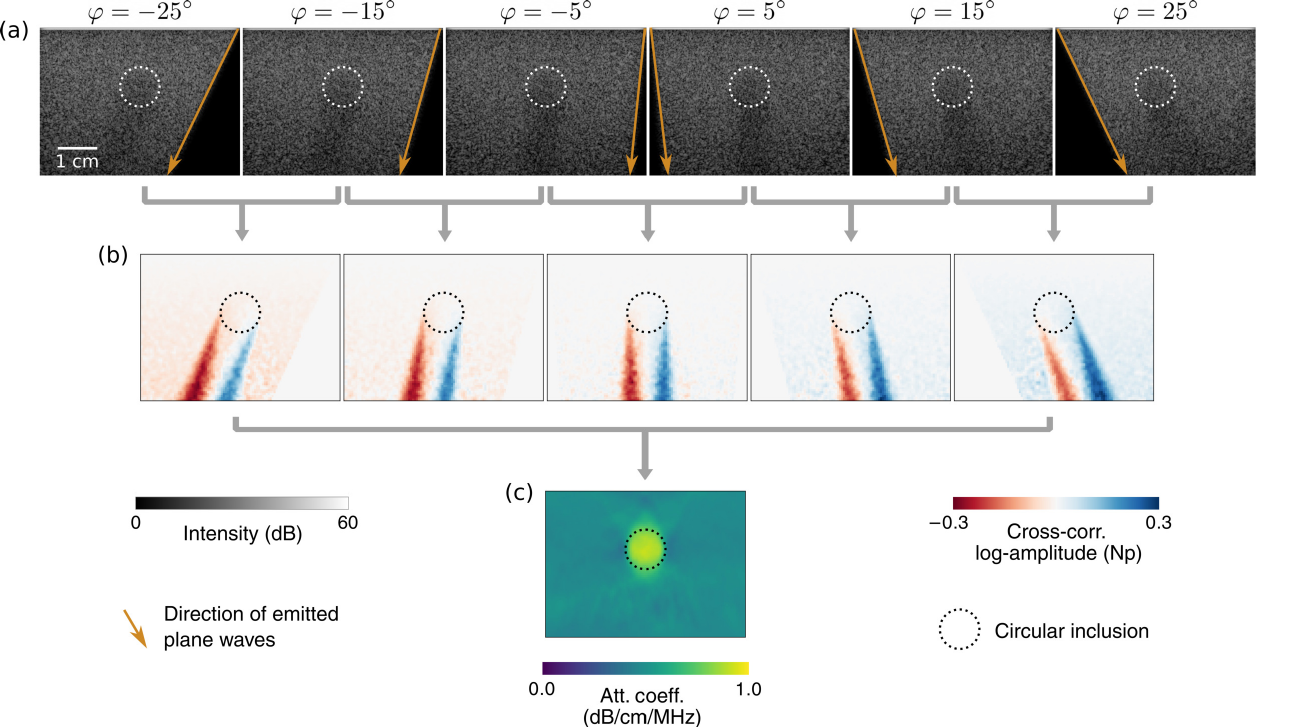}
\caption{Workflow for pulse-echo ultrasound attenuation tomography. (a)~We insonify the tissue by sequentially emitting a set of steered plane waves and reconstruct a complex-valued image per predefined emission angle using delay-and-sum beamforming and coherent compounding. For illustrative purposes, we show B-mode images and increase the angular step to $10^\circ$. (b)~We extract echo log-amplitude changes between emissions by cross-correlating every successive pair of images following equation~\eref{forward_sym}. Nonilluminated areas, shown in white at the margins of the maps, are excluded from the observations. (c)~We solve the regularized linear inverse problem in~\eref{eq:inversion} to retrieve the spatial distribution of attenuation in tissue. The example shown here uses numerically computed ultrasound signals in a medium that contains a circular inclusion with $+0.5~\textrm{dB/cm/MHz}$ attenuation contrast with respect to the background.}
\label{fig:Workflow}
\end{figure}

The quality of the images $I_\varphi(\rvec)$ is degraded due to the clutter arising from, e.g., electronic cross-talk, grating lobes, or multiple scattering that is unaccounted for in the beamforming. To improve this, we apply coherent compounding and collimate the images synthetically on emission along the predefined angles ranging from $-25^\circ$ to $25^\circ$, with an angular step of $2.5^\circ$ (figure~\ref{fig:Workflow}(a)). We use a Gaussian weighting function with the mean equal to the angle of interest and a standard deviation of $3^\circ/\sqrt{2}$. This value is a compromise between reducing clutter and keeping incident beams close to plane waves, as assumed in the forward problem~\eref{forward}.

After coherent compounding, we compute zero-lag cross-correlations between pairs of images of successive angles~\cite{Loupas1995}, and we normalize them with the corresponding auto-correlations following the equation~\eref{forward_sym} (figure~\ref{fig:Workflow}(b)). We use a cross-correlation kernel of 1 mm by 1 mm size around each grid point. While increasing its size reduces noise in log-amplitude measurements, it also reduces the spatial resolution of reconstructed attenuation images. Thus, it must be carefully chosen to optimize this trade-off. 

Due to the limited aperture of the transducer array, the emitted waves do not illuminate the entire image area. We identify the nonilluminated areas by applying geometrical considerations and exclude measurements at these locations from our observations (see white areas in figure~\ref{fig:Workflow}(b)). Finally, we interpolate the measurements into a coarser rectilinear grid with a fixed lateral and axial grid spacing of 0.5 mm to reduce redundancies, and we use the same grid for reconstructing tissue attenuation (figure~\ref{fig:Workflow}(c)).

\subsection{Calibration}

The forward problem in equation \eref{forward} assumes negligible differences between the amplitudes of the cross-correlated plane waves; however, this does not hold in practice due to, e.g., the directivity pattern of transducer elements. To effectively remove transducer-dependent effects, we calibrate our observations with data collected in a reference phantom with known acoustic properties, as is common in state-of-the-art techniques~\cite{Yao90,Labyed2010,Pawlicki2013,Colia2018}. With this calibration, the inverse problem in equation~\eref{eq:inversion} becomes
\begin{equation}
\hat{\mathbf{ m}}  = \mathbf{m}_\mathrm{ref} + \left(\mathbf{F}^\mathsf{T}\mathbf{F} + \lambda \mathbf{D}^\mathsf{T} \mathbf{D} \right)^{-1} \mathbf{F}^\mathsf{T} (\mathbf{ d} - \mathbf{ d_\mathrm{ref}}),
\label{eq:inversionCalibrated}
\end{equation}
where $\mathbf{ d_\mathrm{ref}}$ contains the normalized cross-correlation log-amplitudes measured in the reference phantom with attenuation $\mathbf{m}_\mathrm{ref}$.

\section{Simulation studies}
\label{sec:num_results}

This section uses numerically computed data to  
validate the normalized cross-correlation log-amplitude measurements and discuss the performance of the proposed technique in phantoms with varying contrast in attenuation and echogenicity. We use the k-Wave open-source toolbox~\cite{Treeby2010Kwave} to simulate 2-D ultrasound wave propagation in lossy, nondispersive media. We consider a 256-element transducer array with an inter-element pitch of 0.2 mm, emitting a 7-cycle Gaussian-modulated tone burst with a center frequency of 5 MHz. The media in our examples have a constant density of 1000 kg/m\textsuperscript{2}, a constant speed of sound of 1540 m/s, a homogeneous background with an attenuation coefficient of $\alpha_0 = 0.5~\textrm{dB/cm/MHz}$, and a constant power-law exponent of $y=1$. For calibration data, we use a homogeneous medium with the same speed of sound and $\alpha_0 = 0.2~\textrm{dB/cm/MHz}$. Diffuse scattering is mimicked by introducing random density perturbations with zero-mean Gaussian distribution at each point of the simulation grid, which has a grid spacing of 25~$\mu$m in both lateral and axial directions. All studies in this section use five realizations of the random scattering media and an assumed velocity of 1540~m/s for beamforming. Attenuation reconstructions are obtained with isotropic regularization unless otherwise indicated.

\subsection{Validating normalized cross-correlation log-amplitude measurements}

\begin{figure}[t]
\centering
\includegraphics[width=0.9\textwidth]{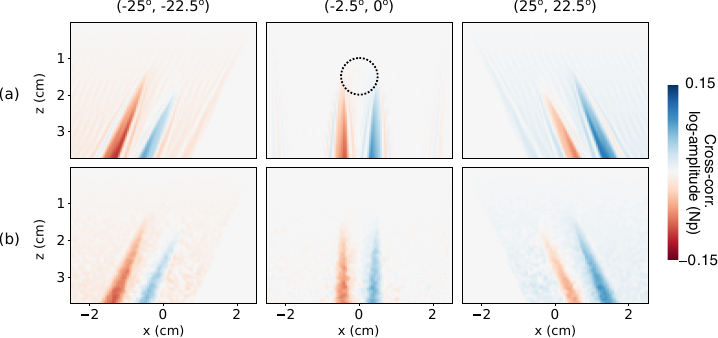}
\caption{Log-amplitude differences between emissions steered at the angles indicated at the top. The medium is homogeneous with a circular inclusion (dashed lines) of positive attenuation contrast. (a) Ground truth. (b) Normalized cross-correlation log-amplitude measurements. Only three exemplary maps out of a total of 20 maps are shown in each case.}
\label{fig:ValLogAmp}
\end{figure}

In~\sref{sec:xcorrtheory}, we showed that the normalized cross-correlation operation is similar to placing virtual receivers inside the medium measuring attenuation-induced differences in the amplitudes of the incident waves. To demonstrate the validity of this conclusion, we consider a medium containing a circular inclusion of ${\alpha_0=1.0~\textrm{dB/cm/MHz}}$ and perform two numerical experiments. In the first experiment, we place the receivers inside the medium at every 12th simulation-grid point (0.3 mm spacing) along both directions. Moreover, we consider plane-wave emissions only at the steering angles of the coherent compounding described in~\sref{sec:measurements}. For each recorded signal, we select the first arrivals and isolate attenuation-related amplitude information using calibration signals. Finally, we compute log-amplitude differences between pairs of emissions of successive angles, shown in figure~\ref{fig:ValLogAmp}(a). We refer to this data as the ground truth. In the second experiment, we consider a pulse-echo ultrasound scenario and follow the approach described in~\sref{sec:implementation} to measure the log-amplitudes of normalized cross-correlations, shown in figure~\ref{fig:ValLogAmp}(b). These measurements capture the ground-truth log-amplitude information remarkably well (mean absolute error (MAE): $3\times10^{-3}$~Np), confirming the accuracy of our theoretical derivations. 

Measured values at each location reflect the differences between the cumulative losses undergone by the incident waves during their propagation. Thus, they are zero at the transducer location and tend to increase with the propagation distance. Below the inclusion, we observe tails of positive and negative values. At these locations, only the propagation path of one of the incident waves crosses the inclusion, leading to large amplitude differences. Interestingly, the ground-truth data shows narrow areas of positive and negative values surrounding each tail, probably caused by the contributions of higher-order Fresnel zones occurring for waves with finite frequencies~\cite{KortaMartiartu2020}. Similar areas are observed for edge waves at the margins of nonilluminated areas. These effects become less noticeable in cross-correlation log-amplitude measurements due to coherent compounding, which acts as a collimator. This will benefit the reconstructions with our proposed technique since the forward modeling in equation~\eref{forward_right} does not account for such wave phenomena. Cross-correlation log-amplitudes also show granular noise that resembles the small-scale fluctuations of tissue echogenicity. This noise is caused by the relatively low number of realizations used for ensemble averaging, which cannot completely remove the imprint of the scatterer distribution. We will discuss the role of this averaging in more detail in section~\ref{sec:ensembleavg} and proceed to analyze the performance of our attenuation imaging technique in the following.

\subsection{Attenuation reconstruction: media with a circular inclusion}

The inverse problem formulated in~\sref{sec:inverse} reconstructs the frequency-dependent quantity $\alpha(\rvec)~=~\alpha_0(\rvec)\omega^{y(\rvec)}$. Yet, throughout this work, we show reconstructed images in terms of the attenuation coefficient $\alpha_0(\rvec)$ to allow direct comparisons with the ground-truth medium properties. We remove the frequency-dependent term from $\alpha(\rvec)$ by using the center frequency of the signals and the known power exponent $y=1$. 

\begin{figure}[t]
\centering
\includegraphics[width=1\textwidth]{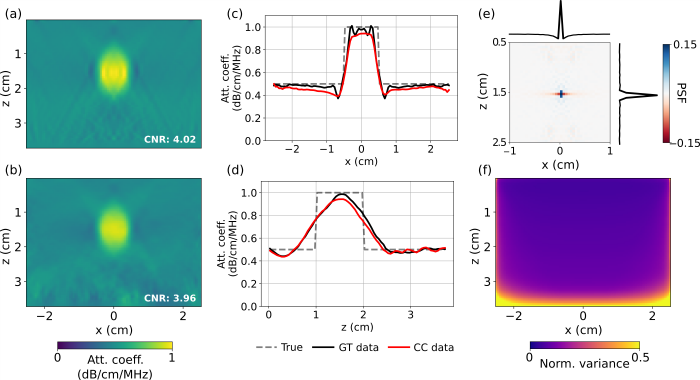}
\caption{Reconstructed attenuation image using (a) the ground-truth (GT) data in figure~\ref{fig:ValLogAmp}(a) and (b) the cross-correlation (CC) data in figure~\ref{fig:ValLogAmp}(b). Both results use the same regularization parameter value $\lambda = 2\times 10^{-6}$, which is optimized for the second reconstruction. The contrast-to-noise ratio (CNR) is computed by taking the values within and outside the true location of the inclusion, covering the whole image area. (c)-(d) Lateral and axial profiles of reconstructed images through the center of the inclusion. Dashed lines show the true attenuation coefficient of the medium. (e) Tomographic point-spread function (PSF) located at the center of the inclusion, together with its lateral and axial profiles. (f) Normalized variance of reconstructed attenuation values, illustrating the relative uncertainties of attenuation estimates at each spatial location.}
\label{fig:Rec&uncertainties}
\end{figure}

First, we want to understand the quality of our reconstructions relative to the highest attainable quality from this type of data. To this end, we compare images retrieved from both the ground-truth data (figure \ref{fig:Rec&uncertainties}(a)) and normalized cross-correlation log-amplitude measurements (figure \ref{fig:Rec&uncertainties}(b)). The corresponding lateral and axial profiles are shown in figures \ref{fig:Rec&uncertainties}(c) and \ref{fig:Rec&uncertainties}(d), respectively. Overall, we observe an excellent agreement between both results, with a mean absolute percentage error (MAPE) between images of 5 \% and the same full width at half maximum (FWHM) in the lateral (0.90 cm) and axial (1.13 cm) profiles. However, the granular noise in cross-correlation data (see figure~\ref{fig:ValLogAmp}) leads to larger errors at the bottom part of the image, slightly diminishing the contrast-to-noise (CNR) ratio. Here, the data coverage is relatively poor, and reconstructed values become highly uncertain (figure \ref{fig:Rec&uncertainties}(f)). In both cases, the reconstructed inclusion appears elongated in the axial direction, showing that our technique provides a lower axial than lateral resolution. This is confirmed by the tomographic point-spread function shown in figure \ref{fig:Rec&uncertainties}(e) and is caused by the limited angular aperture of the steered emissions. Such an aperture moreover leads to X-shaped artifacts crossing the inclusion, also seen in the point-spread function. Since our data cannot constrain the axial resolution well, the regularization enforces smooth features along this direction, spreading the inclusion according to the partial volume effect~\cite{jaeger2022pulseecho}. At the lateral margins of reconstructed inclusions, we observe lower-value areas, which are particularly noticeable for the ground-truth data. There are two factors contributing to this: (i) data from higher-order Fresnel zones, which are neglected by our ray-based forward modeling, and (ii) the relationship between attenuation estimates at different locations shown by the point-spread function in figure \ref{fig:Rec&uncertainties}(e). The latter is intrinsic to our tomographic technique, whereas the first factor is expected to become less relevant at higher frequencies~\cite{KortaMartiartu2020}. This first factor, which is more significant for the ground-truth data (see figure~\ref{fig:ValLogAmp}), is moreover responsible for the line artifacts appearing in figure \ref{fig:Rec&uncertainties}(a).

\begin{figure}[t]
\centering
\includegraphics[width=1.0\textwidth]{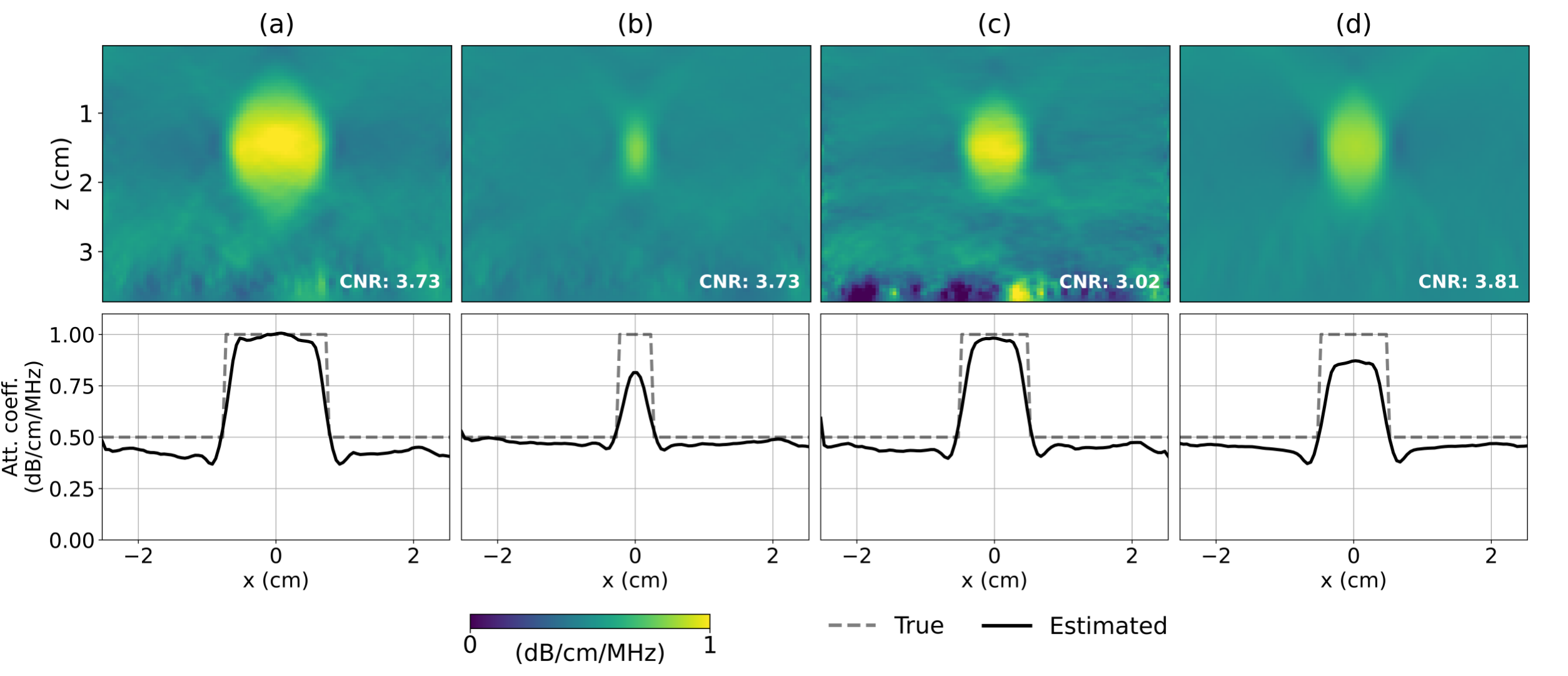}
\caption{Reconstructed attenuation images and their lateral profiles for varying inclusion size and regularization strength. All media have a circular inclusion with $+0.5~\textrm{dB/cm/MHz}$ attenuation contrast. (a), (b) Reconstructions use the same regularization parameter value as in figure~\ref{fig:Rec&uncertainties} ($\lambda=2\times 10^{-6}$), but the circular inclusion has a diameter of ${d=1.5~\textrm{cm}}$ and ${d=0.5~\textrm{cm}}$, respectively. (c), (d) The medium is the same as in figure~\ref{fig:Rec&uncertainties} (${d=1.0~\textrm{cm}}$), but reconstructions use $\lambda = 2\times 10^{-7}$ and $\lambda = 2\times 10^{-5}$, respectively. CNR: Contrast-to-noise ratio.}
\label{fig:RecCircular}
\end{figure}

Reconstructed values within the inclusion are sensitive to the true inclusion size and the regularization. This is shown in figure~\ref{fig:RecCircular}, where we change either the diameter of the circular inclusion or the regularization strength. Compared to the result in figure~\ref{fig:Rec&uncertainties}(b), we observe that the retrieved values within the 1.5-cm-diameter inclusion are closer to their true values (figure~\ref{fig:RecCircular}(a)) but become increasingly underestimated as the inclusion size decreases (figure~\ref{fig:RecCircular}(b)). This is caused by the smoothing effect of the regularization, which reduces the values of structures that are smaller than the resolution limit of our technique. The CNR, however, is the same in figures~\ref{fig:RecCircular}(a) and~\ref{fig:RecCircular}(b) since the former contains more artifacts, especially below the inclusion. Here, the large, highly attenuating area significantly decreases the signal-to-noise ratio of log-amplitude measurements. When we use a weaker regularization than in figure~\ref{fig:Rec&uncertainties}, the attenuation of the inclusion becomes more accurate (figure~\ref{fig:RecCircular}(c)), suggesting that the spatial resolution has improved. Yet, reconstruction artifacts contaminate the image, and the CNR deteriorates. A stronger regularization smooths out the noise at the bottom part of the image but underestimates the attenuation values within the inclusion, having no benefits for the CNR (figure~\ref{fig:RecCircular}(d)). 
The next subsection discusses different regularization strategies particularly suited for layered media.

\subsection{Attenuation reconstruction: layered media}
\label{sec:NumLayered}

\begin{figure}[t]
\centering
\includegraphics[width=0.95\textwidth]{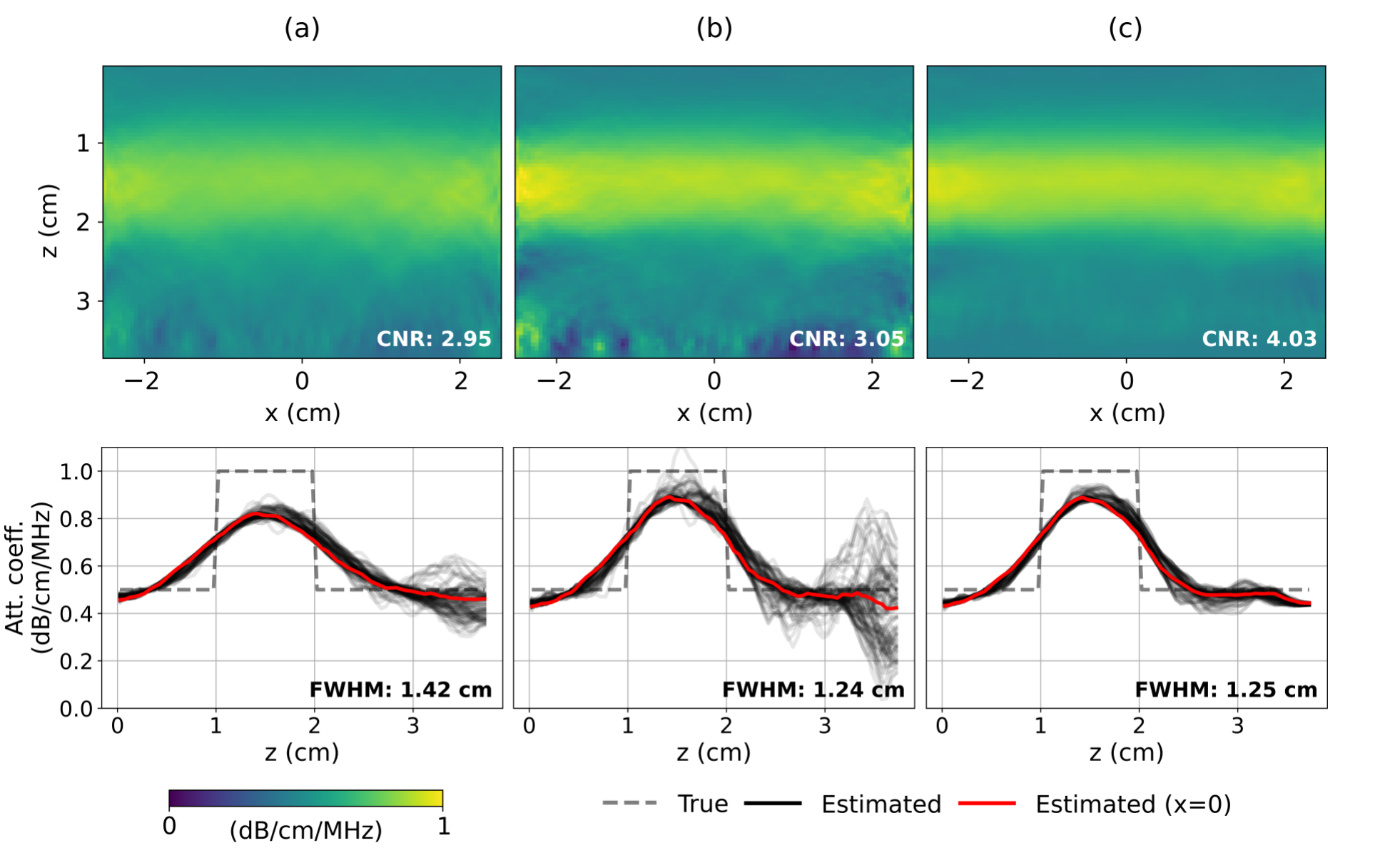}
\caption{Reconstructed attenuation images and their axial profiles for a medium with a 1-cm-thick horizontal layer. Reconstructions use the following regularization parameter values: (a) $\lambda = 2\times10^{-6}$; (b) $\lambda = 4\times10^{-7}$; (c) $\lambda_x = 50\lambda_z$ and $\lambda_z = 4\times10^{-7}$. The profiles show the true attenuation values (dashed gray line) and the estimated ones at every lateral position (solid black line) and position $x=0$ (solid red line). CNR: Contrast-to-noise ratio; FWHM: full width at half maximum. }
\label{fig:RecLayer}
\end{figure}

Previous results demonstrate that our technique provides images with lower axial than lateral resolution. To understand how this affects the retrieval of layered structures, we consider a homogeneous medium containing a 1-cm-thick horizontal layer with $\alpha_0 = 1.0~\textrm{dB/cm/MHz}$. Figure~\ref{fig:RecLayer}(a) shows the reconstructed attenuation image using the same regularization as in figure~\ref{fig:Rec&uncertainties}. Although we recover the layered structure at the correct location, the regularization substantially affects its thickness and attenuation, resulting in a relatively poor CNR. Here, the axial resolution is lower than in figure~\ref{fig:Rec&uncertainties}, where it has probably been enhanced by the lateral variations in the attenuation. As previously discussed, we can improve this with a weaker regularization, shown in figure~\ref{fig:RecLayer}(b). Attenuation estimates within the layer become more accurate, but there is almost no CNR gain due to the strong artifacts appearing in areas with poor data coverage. These examples illustrate the limits of applying the same regularization strength in the axial and lateral directions.

In layered media, we expect smoother features along the lateral than axial direction, and this knowledge can be incorporated into our problem via anisotropic regularization. Figure~\ref{fig:RecLayer}(c) shows the reconstructed attenuation image with the same regularization strength as in figure~\ref{fig:RecLayer}(b) along the axial direction and 50 times stronger along the lateral direction. This approach effectively reduces most artifacts while keeping the axial resolution intact, allowing us to recover the layered structure more accurately and improve the CNR.

\subsection{Attenuation reconstruction: media with echogenicity contrast}

\begin{figure}[t]
\centering
\includegraphics[width=1\textwidth]{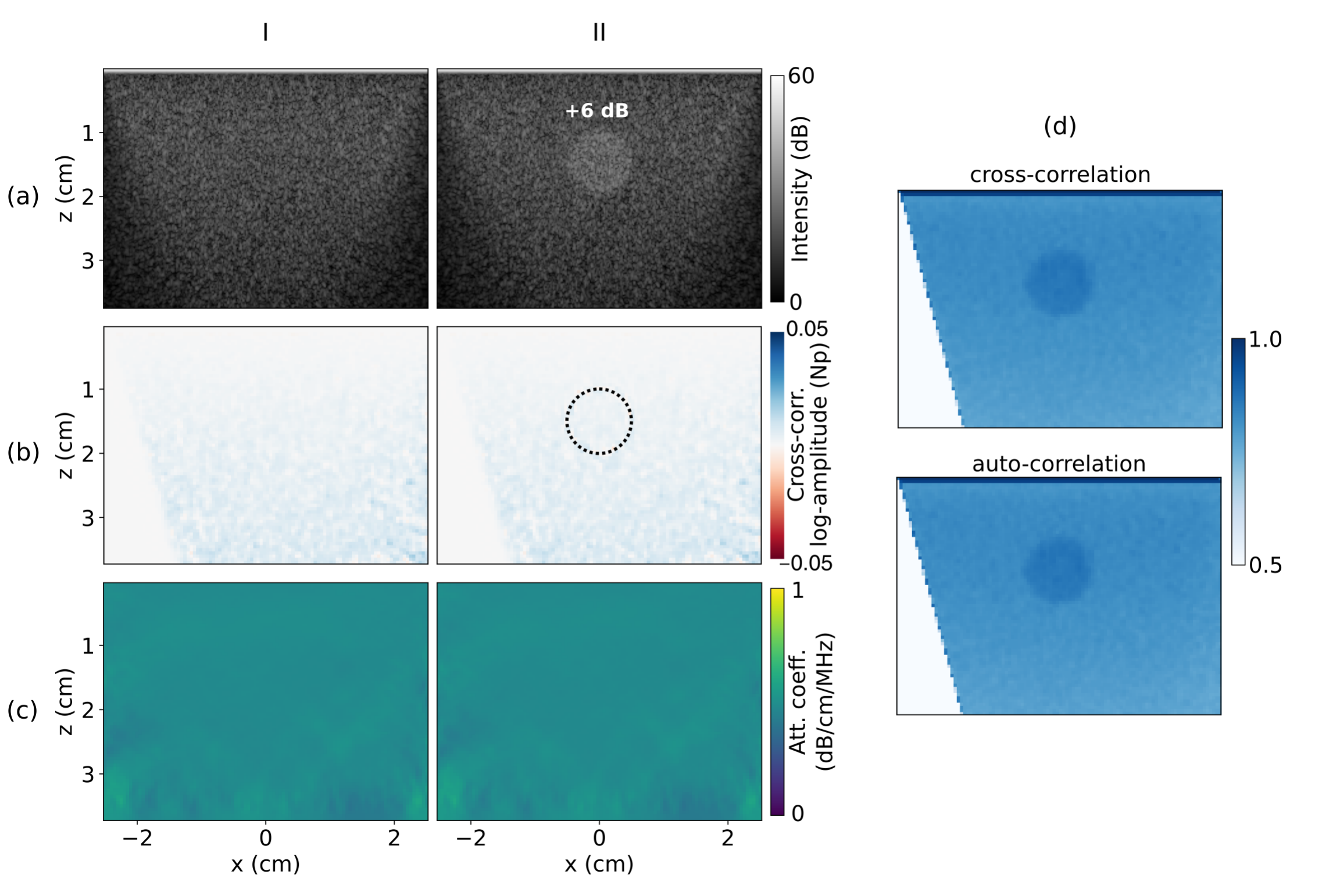}
\caption{(a) B-mode images, (b) an example of normalized cross-correlation log-amplitude measurements (steering angles: 12.5$^\circ$ - 15$^\circ$), and (c) reconstructed attenuation images for a homogeneous medium (I) and the same medium containing a circular inclusion with +6~dB echogenicity contrast (II). Reconstructions use $\lambda = 2\times 10^{-6}$. (d)~Log-amplitudes of cross-correlations and auto-correlations used to compute the measurements in (b) for phantom II. The values are normalized with respect to the maximum log-amplitude in each case.}
\label{fig:RecEcho}
\end{figure}

In section~\ref{sec:xcorrtheory}, we showed that the normalization step is essential to remove the contribution of tissue echogenicity from cross-correlation log-amplitude measurements. To illustrate this better, we consider the two media shown in figure~\ref{fig:RecEcho}(a). Both have the same constant attenuation, but medium II contains a circular inclusion with +6 dB echogenicity contrast. We simulate this by increasing the standard deviation of the Gaussian distribution used to introduce random density perturbations in the medium. Despite the inclusion, normalized cross-correlation log-amplitude measurements are practically identical in both cases (figure~\ref{fig:RecEcho}(b), MAE: $10^{-4}$ Np); thus, the echogenicity contrast has no influence on the images reconstructed with our technique (figure~\ref{fig:RecEcho}(c)). While the amplitudes of cross-correlations do contain variations due to the inclusion (figure~\ref{fig:RecEcho}(d)), the normalization with the auto-correlations effectively removes this information from log-amplitude measurements, as theoretically expected from equation~\eref{forward}. These results are excellent in a context where state-of-the-art spectral-difference methods fail to provide accurate attenuation estimates in tissue with varying echogenicity~\cite{KIM2008,Pawlicki2013}.

\section{Experimental results}
\label{sec:exp_results}

This section aims to demonstrate the experimental feasibility of our attenuation imaging technique using tissue-mimicking phantoms with similar complexity as in previous examples. We also analyze here the influence of incorrect speed-of-sound assumptions on the performance of the technique and discuss the role of ensemble averaging. We acquire ultrasound data using the SuperSonic\textsuperscript{\tiny\textregistered} MACH\textsuperscript{\tiny\textregistered} 30 ultrasound system (Hologic\textsuperscript{\tiny\textregistered} - Supersonic Imagine\textsuperscript{\tiny\textregistered}, Aix en Provence, France) with the L18-5 linear probe. As in our numerical simulations, the probe contains 256 transducer elements with an inter-element pitch of 0.2 mm. Acquired signals have a center frequency of 6.5 MHz with a 0.6 fractional bandwidth and a sampling frequency of 40 MHz. Data is gathered in custom-made calibrated CIRS phantoms (Computerized Imaging Reference Systems, Inc. Norfolk, VA, USA) with inclusions of varying size, echogenicity, and attenuation. Their speed-of-sound values vary less than 5 m/s with respect to the background medium, which has a speed of sound of 1518 m/s. We use this value for delay-and-sum beamforming unless otherwise indicated. We take five measurements per position for the ensemble averaging and ten measurements in a homogeneous area of the phantoms for calibration.

\begin{figure}[t]
\centering
\includegraphics[width=1\textwidth]{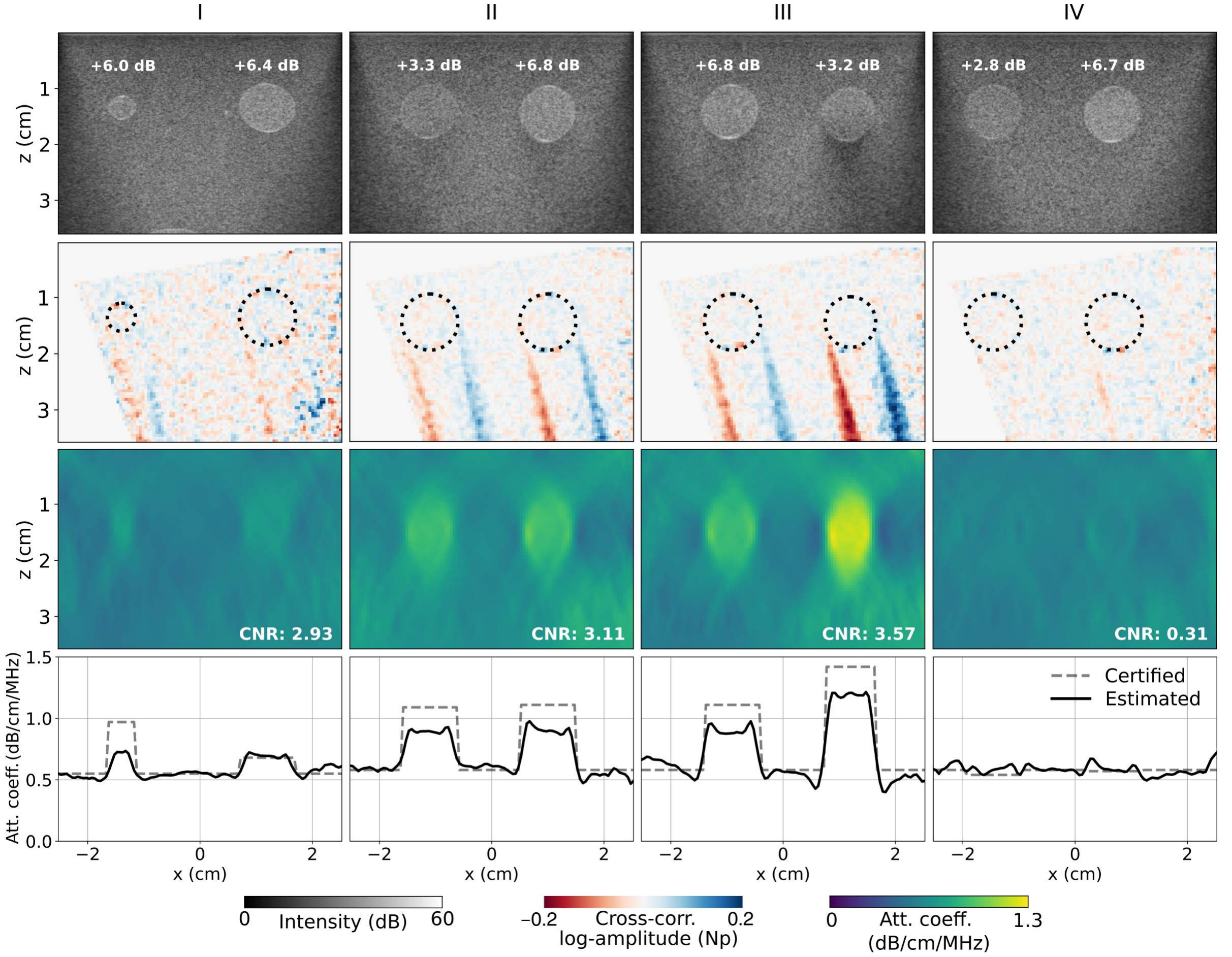}
\caption{Experimental results in tissue-mimicking phantoms containing cylindrical inclusions with contrast in attenuation and echogenicity. The B-mode images in the first row show the shape and location of the inclusions and their echogenicity contrast with respect to the background medium. In the second row, we show an example of normalized cross-correlation log-amplitude measurements (steering angles: 12.5$^\circ$ - 15$^\circ$). Dashed lines indicate the location of the inclusions. The last two rows display the reconstructed attenuation  images, where we indicate the average contrast-to-noise ratio (CNR) of both inclusions, and their lateral profiles through the center of the inclusions. The profiles also show the calibrated attenuation values provided by the phantom manufacturer (dashed gray line). All reconstructions use $\lambda = 3\cdot 10^{-5}$ or $\lambda = 8\cdot 10^{-5}$ depending on the L-curve optimization results (see supplementary figure 1). }
\label{fig:ExpCircular}
\end{figure}

Figure~\ref{fig:ExpCircular} shows the results for phantoms containing cylindrical inclusions. Unlike in our numerical simulations, plane-wave emissions lead to electronic cross-talk noise in the shallow areas, which we exclude to avoid meaningless observations (see the second row in figure~\ref{fig:ExpCircular}). Since our calibration data is acquired in the same phantom, observed experimental data do not contain information about amplitude losses caused by the background attenuating media. We mainly observe inclusion-related tails whose values increase in magnitude with increasing contrast in attenuation. Data noise is larger than in our numerical examples, especially at the margins of the image area, where the focusing quality of beamformed images is lower. Consequently, optimal regularization parameter values are at least one order of magnitude larger than in previous examples. As shown in figure~\ref{fig:RecCircular}(d), this leads to images with a lower axial resolution where inclusions are more axially elongated, reducing their attenuation estimates accordingly. A stronger regularization forces stronger relations between reconstructed values at different locations. As a result, we observe more pronounced X-shaped artifacts centered at the reconstructed inclusions, increasing attenuation estimates of the background medium above and below the inclusions. Similarly, the lower-value areas at the lateral edges of inclusions also become more noticeable, as shown by the reconstruction of, e.g., the largest attenuating inclusion in phantom III. Despite these regularization artifacts, which are also observed in figure~\ref{fig:RecCircular}(d), our results show reasonable average CNR values, consistent with the ground-truth inclusion properties. As in our numerical examples, the relatively good lateral resolution allows us to reconstruct structures that are 0.5 cm small (see phantom I). Moreover, our attenuation imaging technique accurately recovers the relative variations in the attenuation values of the inclusions (e.g., see phantom II and III) and remains insensitive to variations in echogenicity (see phantom IV).

\begin{figure}[t]
\centering
\includegraphics[width=0.9\textwidth]{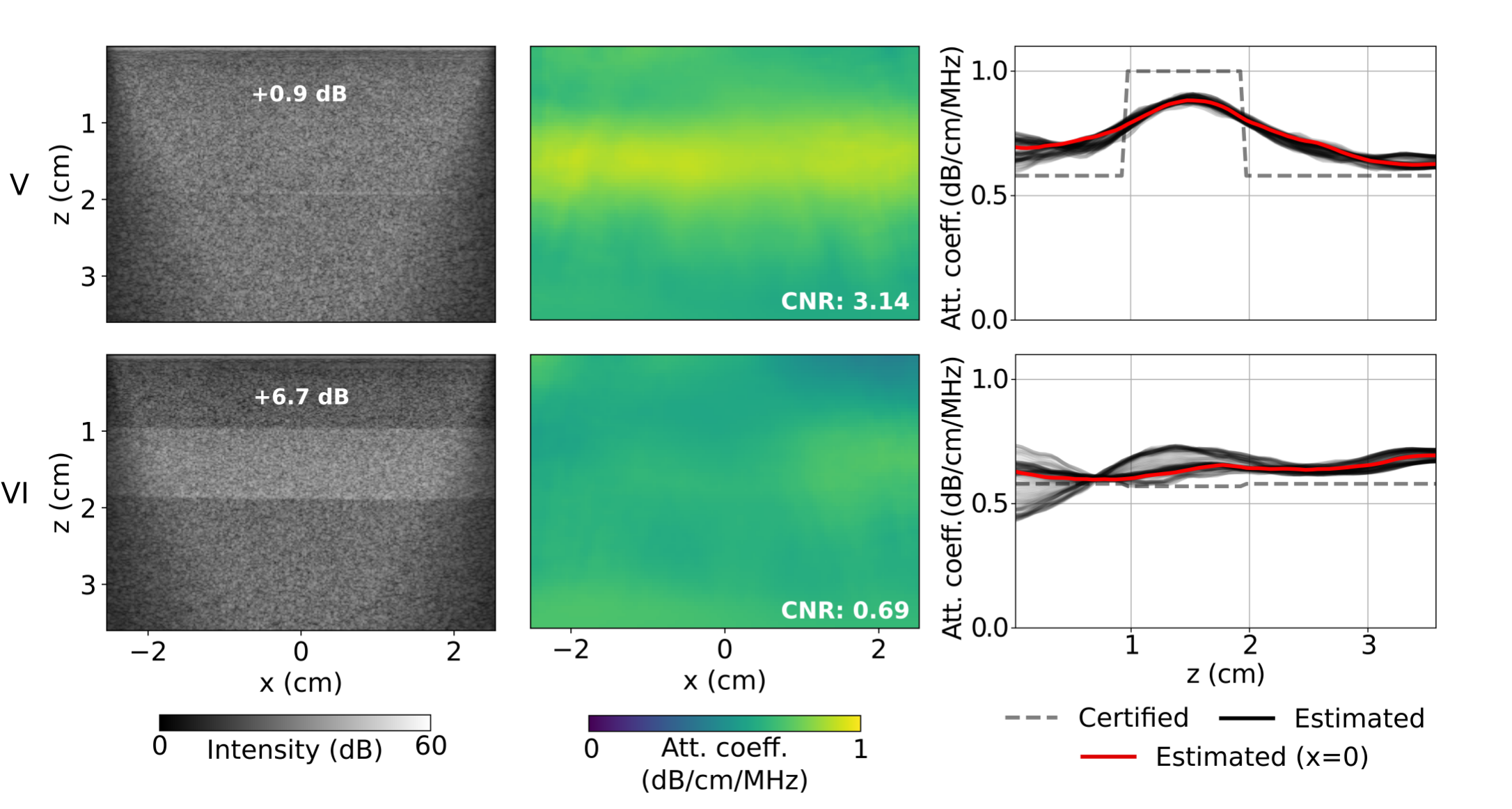}
\caption{Experimental results in tissue-mimicking phantoms containing a horizontal inclusion with contrast in either attenuation (phantom V) or echogenicity (phantom VI). B-mode images in the left column indicate the value of the echogenicity contrast in each case. Reconstructed attenuation coefficient images and their axial profiles are shown in the middle and right columns, respectively. The profiles compare the calibrated attenuation coefficient values provided by the phantom manufacturer (dashed gray line) with the estimated values at each lateral position (solid black line) and position $x=0$ (solid red line). Optimal regularization parameter values are $\lambda_z = 7\cdot 10^{-6}$ and $\lambda_z = 1\cdot 10^{-5}$, respectively. CNR: Contrast-to-noise ratio. }
\label{fig:ExpLayer}
\end{figure}

Similarly, figure~\ref{fig:ExpLayer} shows the experimental results in phantoms containing a horizontal-layer inclusion with contrast in either attenuation or echogenicity. Following our observations in section~\ref{sec:NumLayered}, we use anisotropic regularization satisfying $\lambda_x~=~50\lambda_z$. We keep this $\lambda_x/\lambda_z$ ratio constant in order to optimize the regularization parameter values using the L-curve method (see supplementary figure 2). Although the experimental noise leads to more smoothing than in our numerical examples, and this particularly affects the already limited axial resolution of our technique, we still recover the attenuating inclusion with relatively good CNR and in the correct location (see phantom V). However, the inclusion becomes more diffused in the axial direction, increasing the values of the reconstructed background medium. For phantom VI, our attenuation reconstruction is approximately homogeneous with a variability of 0.05~dB/cm/MHz. There are a few artifacts in the top right part of the image (low and high attenuating areas), but the overall influence of the echogenicity contrast remains relatively low. This example thus confirms the robustness of our technique against variations in echogenicity, as already observed in the previous section.

\subsection{Sensitivity to beamforming speed of sound}
\label{sec:sensiSoS}

\begin{figure}[t]
\centering
\includegraphics[width=1\textwidth]{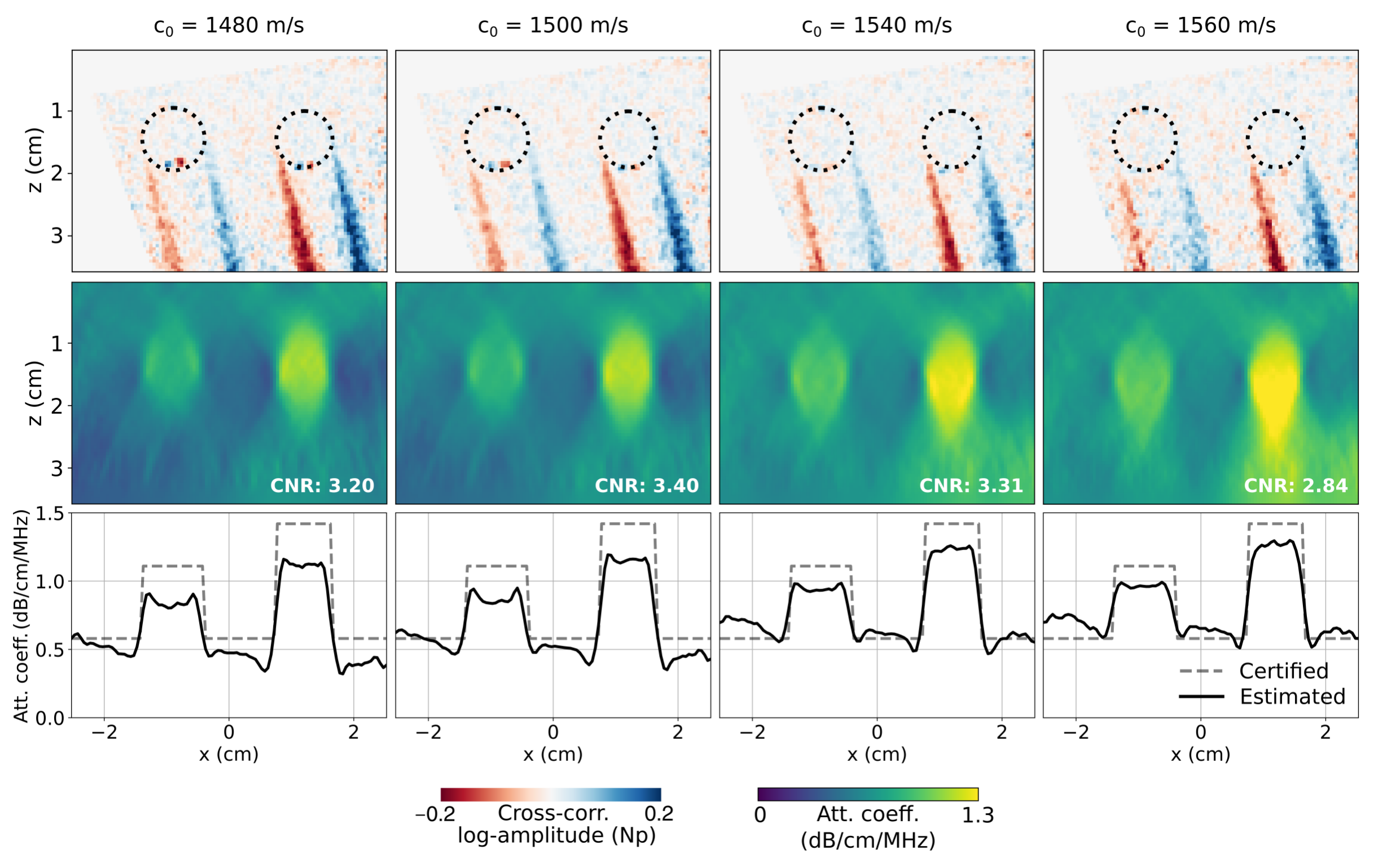}
\caption{Sensitivity of experimental results to the speed-of-sound value $c_0$ used for delay-and-sum beamforming. The first row shows the normalized cross-correlation log-amplitude measurements for steering angles 12.5$^\circ$ - 15$^\circ$. The middle and bottom rows display the reconstructed attenuation images and their lateral profiles, respectively. Considered $c_0$ values in each case are indicated at the top. All reconstructions use $\lambda = 3\cdot 10^{-5}$. CNR: Contrast-to-noise ratio. }
\label{fig:ExpSoSbeam}
\end{figure}

Previous results used the true speed-of-sound value for delay-and-sum beamforming. This ensures optimally focused point-spread functions, allowing us to fulfill the condition~\eref{condition_plane_nearfield2} required for the accuracy of our forward problem. In practice, however, the true tissue speed of sound usually deviates from the assumed speed of sound for beamforming, and aberrations will degrade the focusing quality of beamformed images. To understand how this affects our results, figure~\ref{fig:ExpSoSbeam} compares attenuation images reconstructed with the beamforming speed of sound $c_0$ ranging from 1480~m/s to 1560~m/s. As an example, we consider the phantom III in figure~\ref{fig:ExpCircular}. The results confirm that cross-correlation log-amplitudes measurements become noisier the larger the mismatch between the true and assumed speed of sound is (see the top row in figure~\ref{fig:ExpSoSbeam}). This noise increases with depth as aberrations become stronger and deteriorates the CNR of reconstructed attenuation images. Interestingly, we observe an increase in attenuation estimates with increasing $c_0$. This could indicate that the mislocation of echoes (i.e., our virtual receivers) magnifies the existing gradient with depth in log-amplitude measurements, thereby introducing an apparent shift in attenuation estimates.
Importantly, the reconstructed contrast of inclusions appears insensitive to the choice of $c_0$, meaning that our technique reliably captures the relative attenuation variations of the medium.

\subsection{The role of ensemble averaging}
\label{sec:ensembleavg}

\begin{figure}[t]
\centering
\includegraphics[width=1\textwidth]{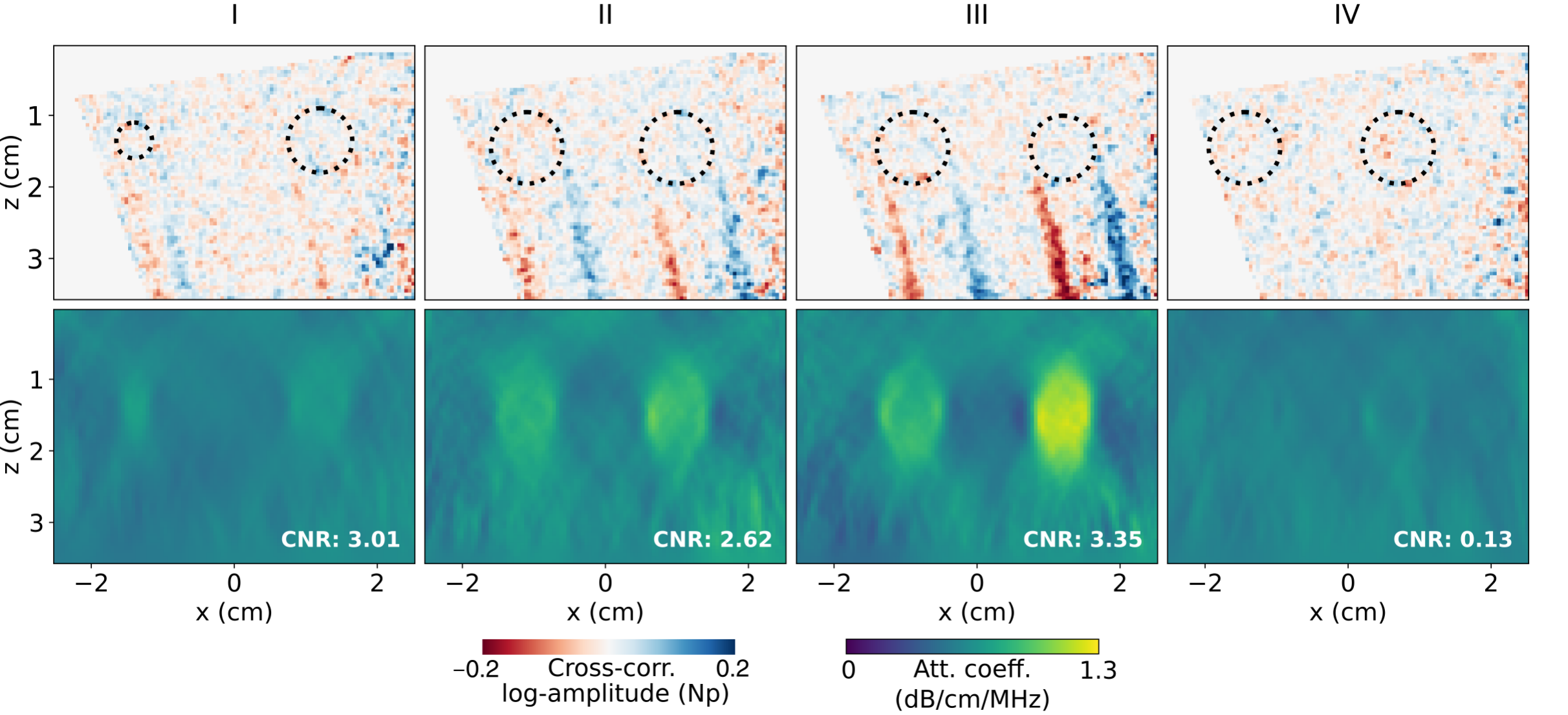}
\caption{Normalized cross-correlation log-amplitude measurements (steering angles: 12.5$^\circ$ - 15$^\circ$) and reconstructed attenuation images for phantoms in figure~\ref{fig:ExpCircular} with only one experimental realization. Dashed lines indicate the location of the inclusions. Reconstructions use the same regularization as in figure~\ref{fig:ExpCircular}. CNR: Contrast-to-noise ratio.}
\label{fig:ExpEnsemble}
\end{figure}

Our forward problem relates the attenuation in tissue to ensemble-averaged cross-correlations. We perform this averaging by taking five repeated measurements at the same location, slightly moving the ultrasound probe. Since this approach can be prone to motion artifacts in vivo, it is important to understand how the ensemble averaging affects the quality of our reconstructions. In figure~\ref{fig:ExpEnsemble}, we show the observed data and reconstructed attenuation images for phantoms I-IV using only one experimental acquisition. Compared to figure~\ref{fig:ExpCircular}, we notice a considerable increase in data noise, particularly at the margins where the focusing quality is the lowest. This noise is random and granular, meaning that the averaging allows us to reduce the imprint of random medium scatterers from the log-amplitude observations. This data noise increases the overall variability of reconstructed attenuation values, which translates to generally lower CNR values, although we still recover the circular inclusions relatively well. Therefore, averaging over repeated acquisitions is not critical to obtain meaningful attenuation images, although it improves their contrast resolution.

\section{Discussion and conclusions}
\label{sec:discussion}

This work presents the first fully 2-D ultrasound attenuation imaging technique for pulse-echo systems. Unlike state-of-the-art techniques, which estimate tissue attenuation from the axial variations of recorded echoes~\cite{Colia2018,Labyed2011}, our approach uses observations of echo-amplitude changes between emissions at different steering angles. We thus constrain the attenuation values at each tissue location using multiple crossing wave paths, essential to resolve the spatial variations of this tissue property~\cite{Rawlinson16}. Furthermore, the proposed approach appears robust against echogenicity contrasts, which are an important source of artifacts for state-of-the-art amplitude-based techniques. These methods neglect the amplitude contributions arising from variations in tissue echogenicity, leading to inaccurate attenuation estimates at the interfaces~\cite{KIM2008}. In contrast, our approach leverages beamformed images obtained for different emissions to effectively remove their imprint from amplitude observations. 

The images reconstructed with our technique show significantly better lateral than axial resolution, which is hampered by the limited angular aperture offered by pulse-echo systems. Current ultrasound probes have an inter-element pitch close to the wavelength of emitted pulses, and grating lobes limit the maximum usable steering angle. The acquisition sequence chosen in this study has been optimized to minimize their influence while maximizing the angular aperture to improve the axial resolution. The small angular step allows us, through the coherent compounding process, to reduce the intensity of echoes arising from grating lobes relative to echoes of the main lobe. The optimal acquisition and processing parameters, however, depend on the respective ultrasound probe in use. High-quality probes featuring a larger element count at a smaller inter-element pitch could allow us to increase the usable angular aperture and thus improve the axial resolution. Regardless of the probe, we could explore other approaches to improve the axial resolution, for instance, by incorporating into our technique observations of axial echo-amplitude changes used in state-of-the-art approaches. 

We base the theoretical derivations of the forward model on two major assumptions. First, we consider tissue as a diffuse scattering medium consisting of randomly and uniformly distributed scatterers. In reality, however, tissue also contains specular reflectors. They reflect ultrasound waves predominantly in directions that depend on the incident steering angle, and the amplitude of echoes detected at these reflectors will vary between different emissions, independently of tissue attenuation. This can introduce outliers in log-amplitude measurements, as we observe at the top and bottom edges of the inclusions in figure~\ref{fig:ExpCircular}. For our experimental examples, only very few measurements showed such values, so they did not significantly affect our results. However, large measurement errors could become important sources of biases for our least-squares-based reconstruction algorithm. It may be interesting to study the performance of our technique with optimization strategies that are more robust to outliers, for instance, by minimizing the $\ell_1$-norm~\cite{Sanabria_2018}.

Second, we have neglected amplitude terms related to the (de)focusing, diffraction, and interference effects caused by tissue speed-of-sound heterogeneities~\cite{Dahlen2002}. If present, they will be mapped as attenuation variations in the reconstructions. Such inaccuracies are not exclusive to our technique but will affect any attenuation imaging method based on amplitude observations. The reconstruction of phantom IV in figure~\ref{fig:ExpCircular} serves as an example to illustrate speed-of-sound-related artifacts. Here, the inclusion on the left has a negligible contrast in attenuation, but cross-correlation phases reveal a significant speed-of-sound contrast with respect to the background (see supplementary figure 3). This is probably why we observe tails in log-amplitude measurements, leading to artifacts at the edges of the inclusion. While these are not concerning in this example, future work should focus on understanding the effects of larger speed-of-sound contrasts and design strategies to disentangle their amplitude contributions from those related to attenuation. 

Our inverse problem is defined to retrieve the frequency-dependent attenuation of tissues. This is a combination of two parameters characterizing tissues: the attenuation coefficient $\alpha_0$ and the frequency power law exponent $y$. In our experimental results, we have assumed a constant value $y=1$ to compare the reconstructed $\alpha_0$ values with those provided by the phantom manufacturer, who reports the averaged measured values in the frequency range 2.25 - 5.5 MHz. The calibrated exponent values of the inclusions, however, vary between 0.86 to 1.17 according to the manufacturer. This may explain part of the mismatch between certified and estimated values observed in figure~\ref{fig:ExpCircular}. To characterize tissue attenuation accurately, both $\alpha_0$ and $y$ should be reconstructed in a multi-parametric fashion. For this, we could adapt our technique, for instance, following the approach suggested by~\citeasnoun{RAU2021} and \citeasnoun{chintada2022spectral}, where we first reconstruct an image per frequency band and then fit a power-law function at every spatial location. In this case, a reduced frequency bandwidth will enlarge the point-spread functions and increase the data noise level, as discussed in~\sref{sec:xcorrtheory}. The feasibility of this approach should therefore be carefully studied in future works.

Similar to other attenuation imaging techniques, calibration is critical to obtain meaningful results with our method. It allows us to minimize amplitude contributions arising from transducer- and system-dependent effects, such as the directivity of transducer elements or time-gain compensation. This approach makes the quantitative accuracy of our images very dependent on a well-calibrated reference phantom  and its acoustic properties, which should be similar to the properties of the tissue under study (e.g., see supplementary figure 4). It would be convenient to reduce this dependency by exploring calibration strategies that use, for example, adjacent frequencies to cancel the unwanted effects~\cite{Gong19}.

The attenuation imaging technique presented in this work relies on similar principles as the speed-of-sound tomography approach CUTE, with comparable computational and technical requirements. One of the main differences between both modalities is the type of observations extracted from the backscattered ultrasound signals: the phases of cross-correlations between beamformed images are related to the speed of sound, whereas their amplitudes are sensitive to attenuation in tissue. Therefore, both imaging modalities are complementary and can be easily combined into a single framework to provide simultaneous images of tissue attenuation and speed of sound. This work thus represents an important step towards exploiting the full waveform content of ultrasound recordings to characterize tissue acoustic properties comprehensively.

\ack

This research has been funded by the Swiss National Science Foundation under the project no. 205320\_179038. N. K. M. thanks the support of the University of Bern through the UniBe Initiator Grant. Calculations were performed on UBELIX (https://www.id.unibe.ch/hpc), the HPC cluster at the University of Bern.

\section*{Data availability statement}
The attenuation imaging Python package developed within this work is available online at the GitHub repository~\url{https://github.com/naiarako/attomo} and Zenodo~\cite{attomo_2023}. The repository also includes the Matlab script used in our examples for generating the numerical wave propagation simulations with k-Wave. The experimental data used in this work can be downloaded via BORIS, the institutional repository of the University of Bern, at \url{https://doi.org/10.48620/390}.
 
\section*{References}
\bibliography{references}

\end{document}